\documentclass[]{spie}  %>>> use for US letter paper
\usepackage{amsmath,amsfonts,amssymb}
\usepackage{graphicx}
\usepackage[colorlinks=true, allcolors=blue]{hyperref}

\usepackage[T1]{fontenc}

\usepackage{adjustbox}
\usepackage{graphicx}
\usepackage{xcolor}
\usepackage{amsmath}
\usepackage{siunitx}
\usepackage{mathtools}
\usepackage{threeparttable}
\usepackage{float}
\usepackage{booktabs}
\usepackage{tabularx}
\usepackage{caption}
\usepackage{subcaption}
\usepackage{multirow}

\title{Longitudinal Assessment of Lung Lesion Burden in CT}

\author[]{Tejas Sudharshan Mathai}
\author[]{Benjamin Hou}
\author[]{Ronald M. Summers}
\affil[]{Radiology and Imaging Sciences, National Institutes of Health (NIH) Clinical Center, USA}

\authorinfo{Send correspondence to T.S.M.: tejas dot mathai at nih dot gov}

\begin{document} 
\maketitle

\begin{abstract}

In the U.S., lung cancer is the second major cause of death. Early detection of suspicious lung nodules is crucial for patient treatment planning, management, and improving outcomes. Many approaches for lung nodule segmentation and volumetric analysis have been proposed, but few have looked at longitudinal changes in total lung tumor burden. In this work, we trained two 3D models (nnUNet) with and without anatomical priors to automatically segment lung lesions and quantified total lesion burden for each patient. The 3D model without priors significantly outperformed ($p < .001$) the model trained with anatomy priors. For detecting clinically significant lesions $>$ 1cm, a precision of 71.3\%, sensitivity of 68.4\%, and F1-score of 69.8\% was achieved. For segmentation, a Dice score of 77.1 $\pm$ 20.3 and Hausdorff distance error of 11.7 $\pm$ 24.1 mm was obtained. The median lesion burden was 6.4 cc (IQR: 2.1, 18.1) and the median volume difference between manual and automated measurements was 0.02 cc (IQR: -2.8, 1.2). Agreements were also evaluated with linear regression and Bland-Altman plots. The proposed approach can produce a personalized evaluation of the total tumor burden for a patient and facilitate interval change tracking over time.

\end{abstract}


%% ===========================
\section{Introduction}
\label{sec_intro}
%% ===========================

Lung cancer is the $2^{nd}$ leading cause of death for men and women in the US with the 5-year survival rate being one of the lowest at 25\% \cite{Siegel2024_lungCancerStats,Blandin2017}. Early detection through lung cancer screening of high-risk patients can lead to improved outcomes \cite{NLST2011,deKoning2020}. Currently, low-dose computed tomography (CT) is preferred for screening patients with lung cancer \cite{Larici12017,Bankier2017} as interval changes in the lesion sizes can be tracked over time \cite{Mozley2012}. A lung nodule is defined as a round or irregular opacity that is either well or poorly defined, measuring $\leq$ 3 cm in diameter, while those $>$ 3 cm are called lung masses \cite{Larici12017,Bankier2017}. According to the Fleischner Society guidelines \cite{Bankier2017}, a strong emphasis is placed on the lung lesion size (mean of long- and short-axis diameters) as an indicator for malignancy \cite{Jaffe2006,Larici12017}. Importantly, lesions in the 3 - 10 mm range have a low cancer risk while those $>$ 1 cm are suspicious for metastasis \cite{Bankier2017}.

% For late stages of lung cancer, the survival rate is even lower at 14–15\%  \cite{Blandin2017}. 

It is cumbersome for a radiologist to manually measure the sizes of multiple lesions in a CT scan during a busy clinical day (40-60 exams), especially with rising CT exam volumes each year \cite{Mahesh2023}. The subjectivity of the sizing measurements and variety of CT scanners and exam protocols also poses a barrier for consistent measurement \cite{Larici12017}. Quantification of total lesion burden is especially useful in patients with advanced lung cancer undergoing Targeted Radionuclide Therapy (TRT). \cite{Leung2024,Jennings2004,Petrick2014} These patients have multiple staging CT and PET-CT studies acquired for theranostics \cite{Lenz2023} and pharmacological treatment \cite{Tricarico2024}. Several prior works on automated lung lesion segmentation for volumetric analysis have been proposed \cite{Liu2022,Mellon2023,Yoon2024,Qi2020,Xu2021,Peters2023,Li2019,Hering2021}. However, most prior approaches relied on proprietary datasets that are not publicly available or used commercially-developed algorithms.  

% Huang2020,Tao2022,Borghesi2023

In this work, lung lesions were segmented in longitudinal CT studies and the total lesion burden was quantified for interval change assessment. The public UniToChest dataset \cite{Chaudhry2022} was used to train a 3D segmentation model (full-resolution nnUNet) to segment lung lesions. It was tested on patients with longitudinal CT exams from the same dataset. Then, interval volumetric changes were determined between the initial and follow-up studies of a patient, thereby providing a personalized estimate of lesion burden. The effect of using anatomical priors from a publicly available tool (TotalSegmentator) \cite{Wasserthal2023_TS} for improvements in lung lesion segmentation was also assessed. The novel contribution is in the longitudinal assessment of lung lesion burden over time for interval change tracking using off-the-shelf segmentation models. 

%%%%%%%%%%%%%%%%%%%%%%%%%%%%%%%%%%%%%%%%%%%%%%%%%%%%%%%%%%%%%%%%%%
\begin{figure*}[!tb]
\centering
\includegraphics[width=\textwidth]{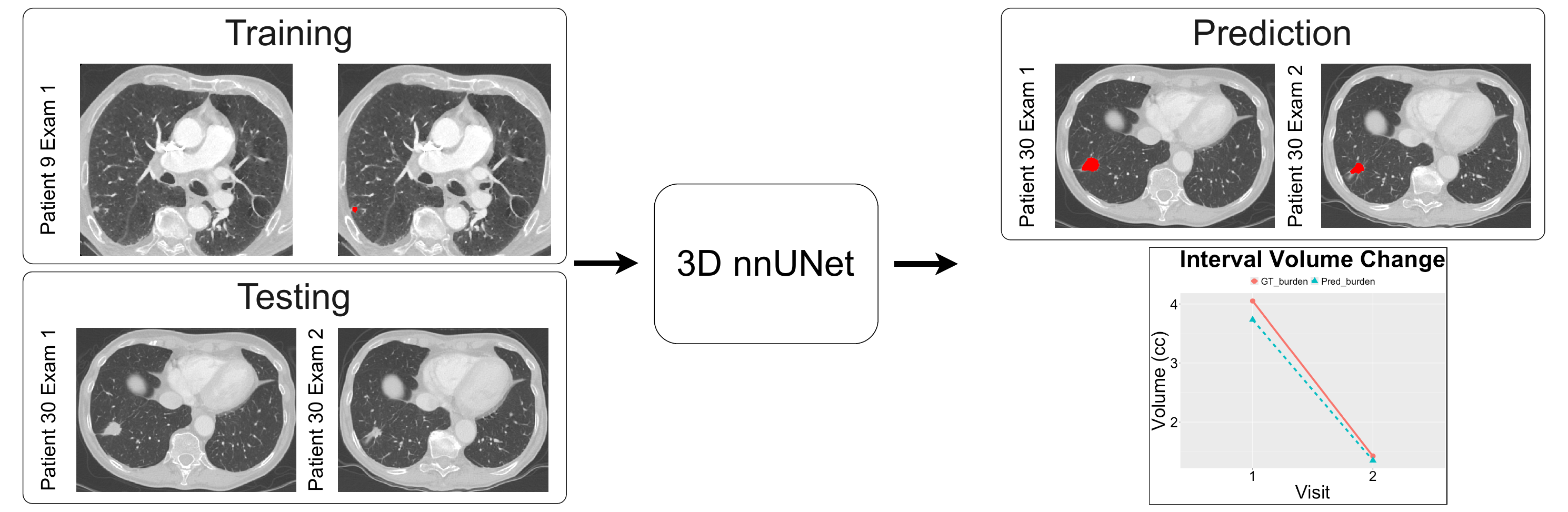}
\caption{Flowchart of the lung lesion segmentation pipeline and assessment of interval changes in total lesion burden. A 3D full-resolution nnUNet was trained to segment lung lesions (red) in patients with longitudinal exams from the public UniToChest dataset. As shown in the graph, a personalized estimate of the total lung lesion burden for a patient (ID \#30) was obtained for interval change tracking over time. }
\label{fig_money}
\end{figure*}
%%%%%%%%%%%%%%%%%%%%%%%%%%%%%%%%%%%%%%%%%%%%%%%%%%%%%%%%%%%%%%%%%%

%% ===========================
\section{Methods}
\label{sec_methods}
%% ===========================

\textbf{Patient Sample.} The public UniToChest dataset \cite{Chaudhry2022} was used in this work. It was acquired at the Citta della Salute e della Scienze Hospital in Italy and comprised of 715 contrast-enhanced thoracic CT volumes from 623 patients. The patient cohort (ages: 17 - 90 years) included 377 males and 246 females who were imaged with 10 different CT scanners (GE, Philips, Siemens). The full 3D extents of 10,071 nodules were manually annotated by a radiologist (unknown years of experience) in the 715 volumes. The CT volume dimensions ranged from 512 $\times$ 512 $\times$ (31 $\sim$ 1135) voxels, while the voxel spacing ranged from (0.41 $\sim$ 1.17) $\times$ (0.41 $\sim$ 1.17) $\times$ (0.24 $\sim$ 5) mm. Amongst the 623 patients, 546 underwent exactly 1 CT exam (7084 nodules) and were solely used for training the segmentation model. The remaining 77 patients had 2+ CT exams (169 studies, 3297 nodules); 62 patients had 2 visits and 15 patients had 3 visits. Micro-nodules ($<$ 3mm, n = 42) were excluded, and the patients with multiple visits were placed in the test subset for longitudinal lesion burden assessment. 

% To the best of our knowledge, there are no publicly available datasets with multiple patient visits except for the UniToChest dataset. However, this dataset was originally divided into train/val/test splits where studies from the same patient could be in more than one split (data leakage). Therefore, all patients with multiple visits were extracted and placed in just the test split in our work

\noindent
\textbf{Model.} Fig. \ref{fig_money} shows an overview of the framework. The self-configuring nnUNet segmentation framework \cite{Isensee_2020} was utilized to segment lung lesions. It is the \textit{de-facto} standard for segmentation tasks due to its award-winning performance \cite{Isensee_2020,Isensee2024} and rigorous validation \cite{Isensee2024} for many tasks, including multi-organ segmentation in CT and MRI \cite{Zhuang2024,Wasserthal2023_TS} among others. It has often outperformed other architectures \cite{Isensee2024}, such as transformer-based approaches \cite{Zhou2023_nnFormer}.  The framework automatically determines a ``fingerprint'' for the input CT volumes that includes intensity normalization and resampling to a consistent spacing among others. The model learned to segment the targets in the CT volume and iteratively refined it via a loss function, which was a combination of the Dice loss and the binary cross entropy loss \cite{drozdzal2016importance}. Additional details on implementation are provided in the Appendix. 

\noindent
\textbf{Experiments.} In the first experiment, a 3D full-resolution nnUNet model was trained directly with the ground-truth lesions provided in the public UniToChest dataset. This model is called ``\texttt{noPriors}'' henceforth. The second experiment was to evaluate the use of 28 anatomical priors \cite{Mathai2024} derived from the public TotalSegmentator tool \cite{Wasserthal2023_TS}. Another 3D full-resolution nnUNet model was trained with the 28 anatomy priors and lesion labels (total of 29 labels), and this model is called ``\texttt{withPriors}'' going forward. Lung lesions can often be identified at the apex of a lobe and at transitional regions between the lung and adjacent structures (e.g., airways, pleura, diaphragm). The purpose of the second experiment was to assess the benefit of the anatomy priors in differentiating between adjacent structures with similar intensities \cite{Bouget2022_StOlavs}, while simultaneously combating class imbalance (background vs. lesions) during model training. 

\noindent
\textbf{Statistical Analysis.} Detection performance was measured through precision, sensitivity, and F1 scores. For segmentation performance, the Dice Similarity Coefficient (DSC) and symmetric Hausdorff Distance (HD) were calculated. A non-parametric Wilcoxon signed-rank test was performed to assess the utility of anatomical priors for lesion segmentation. A $p < 0.05$ indicated statistical significance. Additional statistical measures (linear regression and Bland-Altman plots) were also computed and are described in the Appendix. 

%% ===========================
\section{Results}
\label{sec_results}
%% ===========================

\begin{figure}[!t]
    \centering
    \begin{subfigure}[b]{0.325\textwidth}
        \centering
        \includegraphics[width=\textwidth]{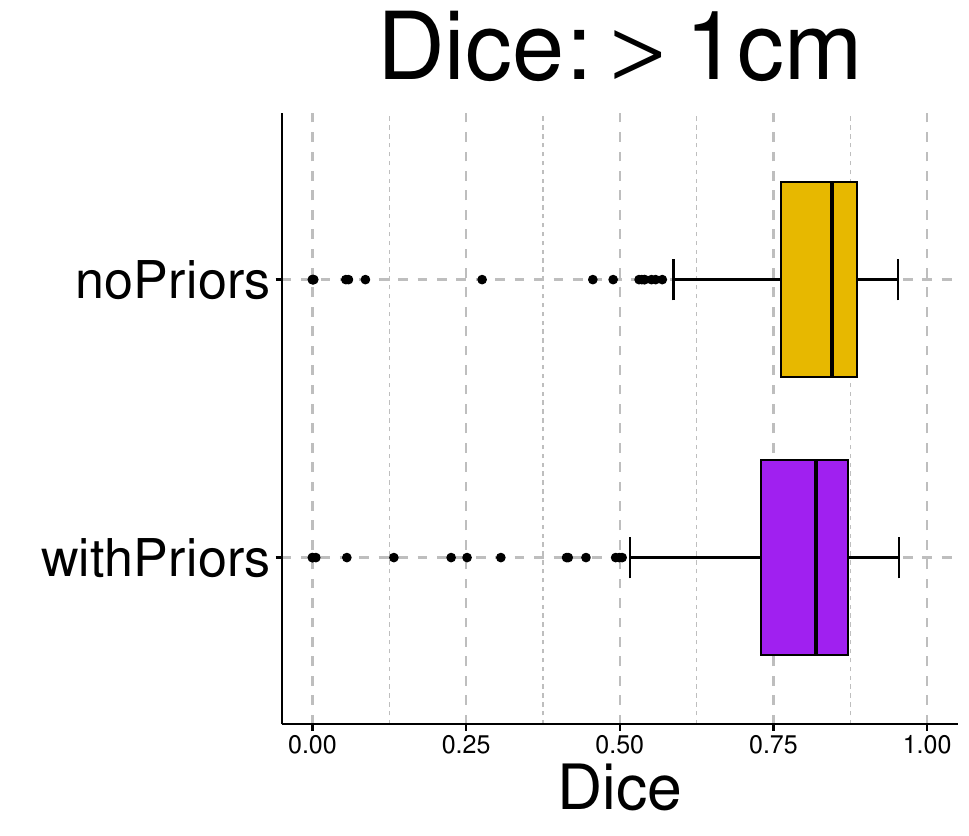}
        \caption{\small Dice: lesions $>$ 1cm}
    \end{subfigure}
    \hfill
    \begin{subfigure}[b]{0.325\textwidth}  
        \centering 
        \includegraphics[width=\textwidth]{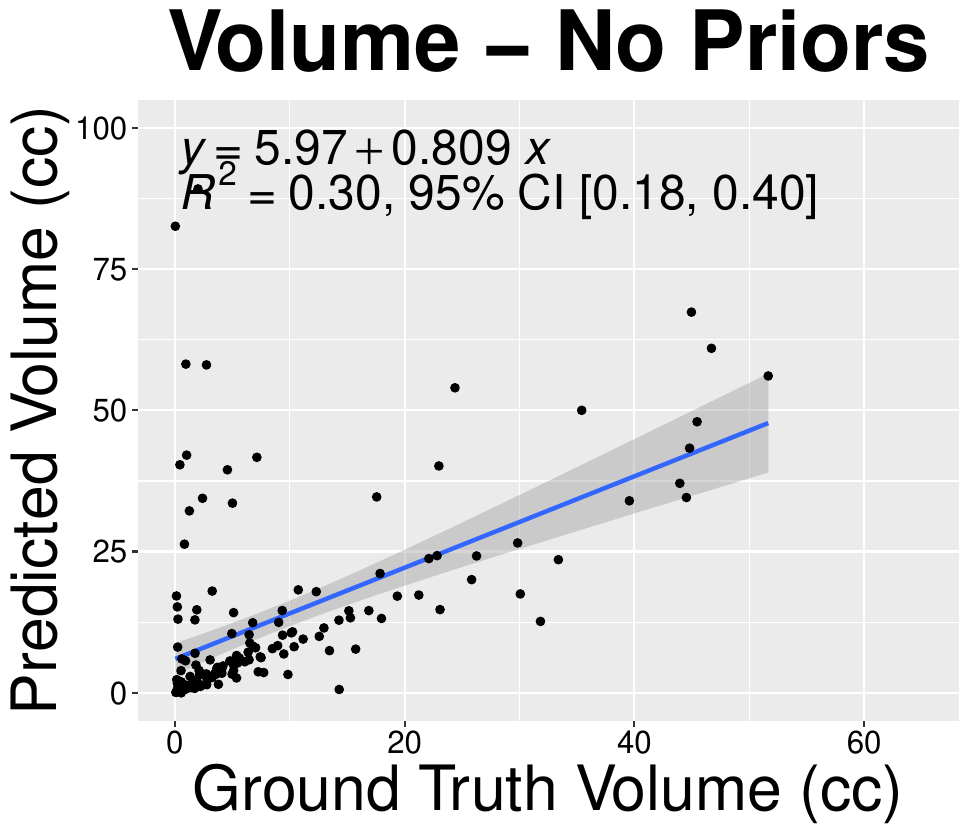}
        \caption{\small ${R}^{2}$, no priors}
    \end{subfigure}
    \hfill
    \begin{subfigure}[b]{0.325\textwidth}  
        \centering 
        \includegraphics[width=\textwidth]{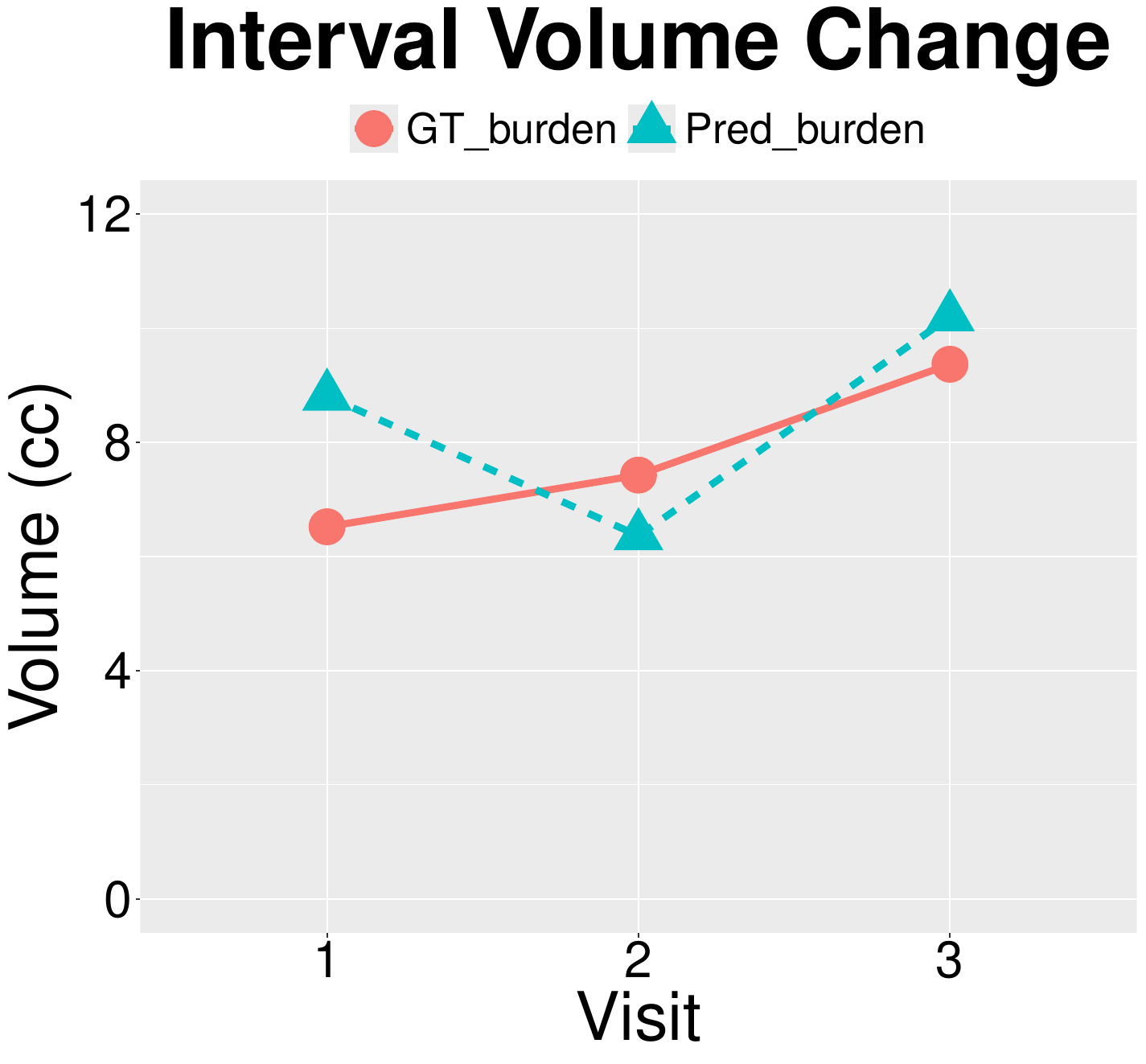}
        \caption{\small Patient 533}
    \end{subfigure}
    \vskip\baselineskip
    \begin{subfigure}[b]{0.325\textwidth}   
        \centering 
        \includegraphics[width=\textwidth]{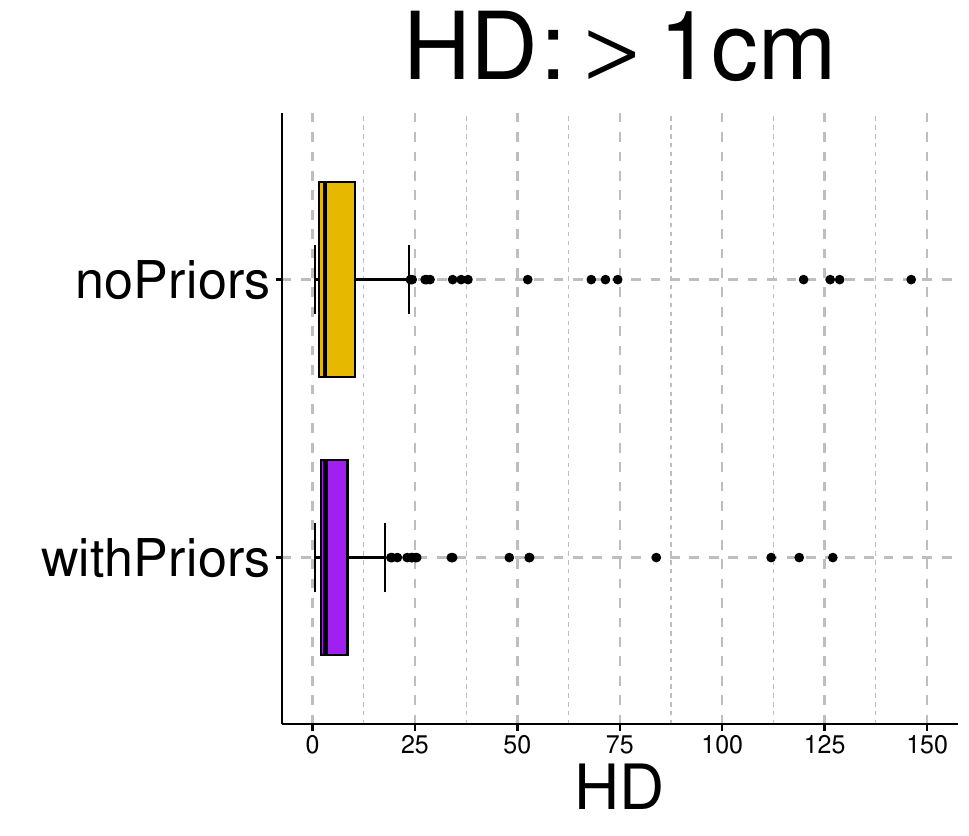}
        \caption{\small HD: lesions $>$ 1cm}
    \end{subfigure}
    \hfill
    \begin{subfigure}[b]{0.325\textwidth}   
        \centering 
        \includegraphics[width=\textwidth]{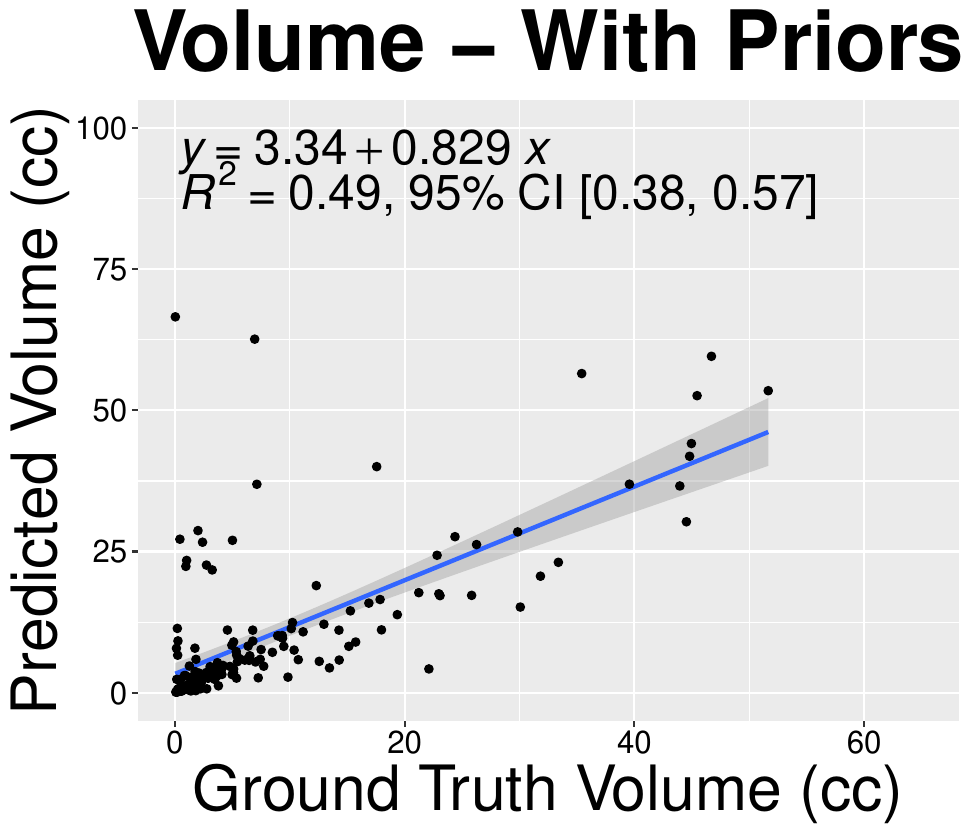}
        \caption{\small ${R}^{2}$, with priors}
    \end{subfigure}
    \hfill
    \begin{subfigure}[b]{0.325\textwidth}   
        \centering 
        \includegraphics[width=\textwidth]{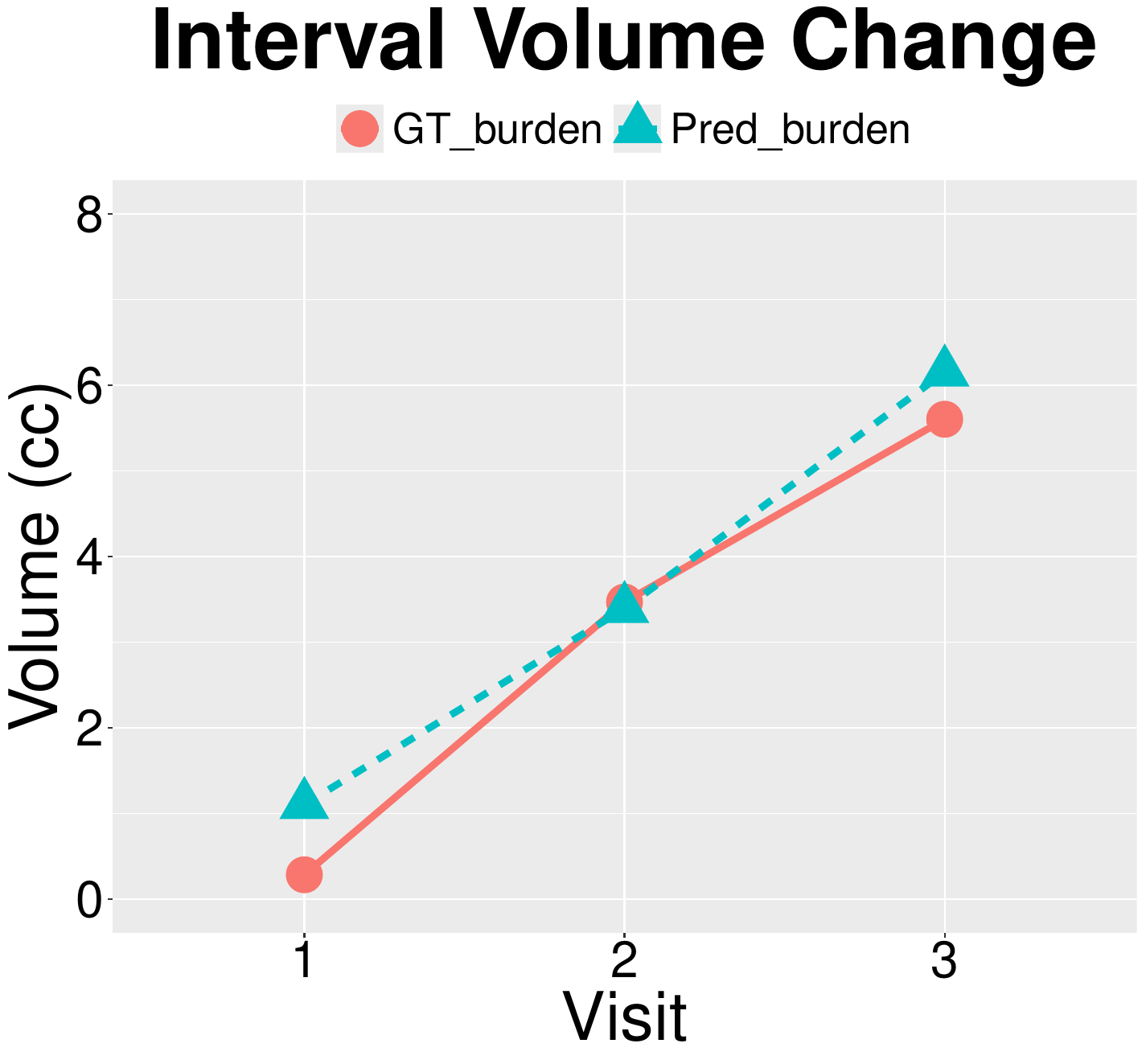}
        \caption{\small Patient 606}
    \end{subfigure}
    \smallskip
    \caption{Dice scores (a) and Hausdorff distance (HD) errors (d) for lung lesions $>$ 1cm. Correlation of reference vs. predicted lesion volumes for lesions segmented by 3D nnUNet model without (b) and with anatomy priors (e) for all test CT data. Total lung lesion burden over time for two patients (c, f) as computed by 3D nnUNet trained without priors.} 
    \label{fig_results_bxp_r2_traj}
\end{figure}

Table \ref{table_detectionResults} presents the results of lesion detection, while Table \ref{table_segmentationResults} summarizes the lesion segmentation performance. Fig. \ref{fig_results_bxp_r2_traj} shows results of different segmentation metrics. For lesion detection, the 3D nnUNet model (``\texttt{noPriors}'') trained without priors fared better than the model ``\texttt{withPriors}'' across all metrics with the exception of a lower precision for lesions $>$ 1 cm. For lesion segmentation, the ``\texttt{noPriors}'' model attained higher Dice scores across all lesion sizes. The Wilcoxon signed-rank test revealed a significant difference between the Dice scores of 3D ``\texttt{noPriors}'' model vs. ``\texttt{withPriors}'' model for all lung lesions ($p < .001$, moderate effect size r =.43), lesions 3 - 10 mm ($p < .001$, r =.39), and clinically relevant lesions $\geq$ 1 cm ($p < .001$, r =.39). There were no significant differences between the HD errors for the ``\texttt{noPriors}'' vs. ``\texttt{withPriors}'' model across the three categories. 

The lower HD errors for the ``\texttt{withPriors}'' model could be due to the penalty imposed by the model on predictions that encroached into the adjacent anatomy (lung lobes, arteries, veins etc.). The ``\texttt{noPriors}'' model saw an increase in HD errors with the removal of the penalty. A Friedman test yielded a $p = 0.58$, which indicated that the median of the Dice score distributions were not different across patient studies, indicating that the models were not biased to the first exam. A greater agreement between the ground-truth and predicted lesions was seen for the ``\texttt{withPriors}'' model compared to the ``\texttt{noPriors}'' model ($R^{2} = .49$ vs. $R^{2} = .3$). For the ``\texttt{noPriors}'' and ``\texttt{withPriors}'' models, the median volume difference (signed) between the manual and automated measurements were 0.02 cc (IQR: -2.8, 1.2) and -0.003 cc (IQR: -1.4, 1.4), respectively. Additional figures and summary results are presented in the Appendix.

% Fig. \ref{fig_example_CT_predictions} shows a few example outputs from the models. 

% supplementary
% \ref{fig_results_boxplots}

%%%%%%%%%%%%%%%%%%%%%%%%%%%%%%%%%%%%%%%%%%%%%%%%%%%%%%%%%%%%%%%%%%
%%%%%%%%%%%%%%%%%%%%%%%%%%%%%%%%%%%%%%%%%%%%%%%%%%%%%%%%%%%%%%%%%%

%%%%%%%%%%%%%%%%%%%%%%%%%%%%%%%%%%%%%%%%%%%%%%%%%%%%%%%%%%%%%%%%%%
\begin{table}[!h]
\centering
\caption{Results of lung lesion detection for different sizes. Bold font indicates best results.}
\begin{adjustbox}{max width=\textwidth}
\begin{tabular}{@{} l c *{9}{c} @{}}
  \toprule
  Dataset & Priors &
    \multicolumn{3}{c}{Precision $\uparrow$} &
    \multicolumn{3}{c}{Sensitivity $\uparrow$} &
    \multicolumn{3}{c@{}}{F1 score $\uparrow$} \\
  \cmidrule(lr){3-5} \cmidrule(lr){6-8} \cmidrule(l){9-11}
  & & All & 3 - 10 mm & $>$ 1 cm & All & 3 - 10 mm & $>$ 1 cm & All & 3 - 10 mm & $>$ 1 cm \\
  \midrule
  3D nnUNet & No     & \textbf{65.6} & \textbf{57.6} & 71.3 & \textbf{49.9} & \textbf{42.5} & \textbf{68.4} & \textbf{56.7} & \textbf{48.9} & \textbf{69.8} \\
  3D nnUNet & Yes    & 62.1 & 51.7 & \textbf{75.1} & 46.2 & 38.1 & 64.9 & 52.9 & 43.9 & 69.7 \\
  \bottomrule
\end{tabular}
\end{adjustbox}
\label{table_detectionResults}
\end{table}
%%%%%%%%%%%%%%%%%%%%%%%%%%%%%%%%%%%%%%%%%%%%%%%%%%%%%%%%%%%%%%%%%%

%%%%%%%%%%%%%%%%%%%%%%%%%%%%%%%%%%%%%%%%%%%%%%%%%%%%%%%%%%%%%%%%%%
\begin{table*}[!h]
\centering
\caption{Results of segmenting lung lesions of different sizes. Standard deviations are shown in parentheses. Bold font indicates best results.}
\begin{adjustbox}{max width=\textwidth}
\begin{tabular}{l @{\quad} c @{\quad} ccc @{\quad} ccc}
  \toprule
  \multirow{2}{*}{\raisebox{-\heavyrulewidth}{Experiment}} & \multirow{2}{*}{\raisebox{-\heavyrulewidth}{Priors}} & \multicolumn{3}{c}{Dice $\uparrow$} & \multicolumn{3}{c}{Hausdorff Distance $\downarrow$} \\
  \cmidrule{3-8}
  & & All & 3 - 10 mm & $>$ 1 cm & All & 3 - 10 mm & $>$ 1 cm \\
  \midrule
  3D nnUNet & No     & \textbf{74.5 (19.1)} & \textbf{52.5 (26.9)} & \textbf{77.1 (20.3)} & 21.3 (42.1) & 40.5 (60.8) & 11.7 (24.1) \\
  3D nnUNet & Yes    & 72.8 (18.1) & 49.7 (26.1) & 75.6 (19.2) & \textbf{19.2 (34.2)} & \textbf{34.9 (48.5)} & \textbf{9.8 (19.7)} \\
  \bottomrule
\end{tabular}
\end{adjustbox}
\label{table_segmentationResults}
\end{table*}
%%%%%%%%%%%%%%%%%%%%%%%%%%%%%%%%%%%%%%%%%%%%%%%%%%%%%%%%%%%%%%%%%%

%% ===========================
\section{Discussion}
\label{sec_discussion}
%% ===========================

In this work, two 3D full-resolution nnUNet models (with and without anatomical priors) were trained to segment lung lesions in CT volumes of patients with single visits, and then evaluated on longitudinal exams. As shown in Figs. \ref{fig_money} and \ref{fig_results_bxp_r2_traj}, a personalized assessment of total lesion burden was computed for patients with longitudinal visits to track interval changes over time. In terms of clinical applicability, the results from this work are similar to that obtained in prior work \cite{Kashyap2025}. The proposed pipeline is fully automated, trained on a public dataset, and required no manual intervention. The ``\texttt{noPriors}'' model fared the best for lesions of all sizes in terms of detection and segmentation performance. The approach can analyze interval changes in either the left or right lung alone, and in each pulmonary segment. It shows promise for lesion burden tracking in patients with a primary lung cancer diagnosis, lymph node involvement \cite{Taupitz2007} and lung metastasis from non-lung cancers \cite{Jha2024}.

A quality check (qualitative) of the segmentation labels provided with the public UniToChest dataset was conducted, and issues with label noise in the dataset were identified. First, small clusters of voxels that did not belong to the main lesion (main connected component) were seen. Second, several lesions had a few of the slices entirely unlabeled, and this effectively divided a single lesion into multiple pieces. The division not only affected the detection and segmentation performance, but also the total lesion burden estimation and comparison over time. This is the main limitation of the work. It is hypothesized that once these label errors are corrected, the segmentation performance and the volumetric agreements could improve for both models. Interestingly, the ``\texttt{noPriors}'' model generated false positives that fell outside the lung region, whereas the ``\texttt{withPriors}'' model yielded segmentations that were only within the lungs. The lack of anatomy priors did not constrain the ``\texttt{noPriors}'' model predictions to just the lung region, and indicates a benefit to using anatomy priors despite its lower performance \cite{Mathai2024,Engelson2024}. Another limitation of this work was the lack of an association with a clinical disease correlate (e.g, non-small cell lung cancer); the public UniToChest dataset did not contain underlying disease-specific information.

For irregular tumor growth, there is subjectivity in the linear measurement of size (with electronic calipers) by radiologists during a busy clinical day \cite{Larici12017}. To ameliorate this issue, automated calculation of volumetric measurements holds promise for long-term patient care. Additionally, it can flag any concerning lesions that were missed by the reading radiologist during a ``second'' automated read. But, the integration of such tools into the clinical workflow has seen a slow uptake due to several reasons \cite{Mellon2023}. Incorporation of such a tool into the Picture Archival and Communication System (PACS) at hospitals can enable radiologists to move beyond image interpretation and report writing to higher-value tasks, such as overall clinical management of patients \cite{Mellon2023}. 

\clearpage

\section{Acknowledgements} 

This work was supported by the Intramural Research Program of the NIH Clinical Center (project number 1Z01 CL040004).

%% References
\bibliography{references} % bibliography data in report.bib

\begin{thebibliography}{10}

\bibitem{Siegel2024_lungCancerStats}
Siegel, R.~L., Giaquinto, A.~N., and Jemal, A., ``Cancer statistics, 2024,'' {\em CA: A Cancer Journal for Clinicians}~{\bf 74}(1),  12--49 (2024).

\bibitem{Blandin2017}
Blandin~Knight, S. et~al., ``Progress and prospects of early detection in lung cancer,'' {\em Open Biology}~{\bf 7}(9),  170070 (2017).

\bibitem{NLST2011}
Team, T. N. L. S. T.~R., ``Reduced lung-cancer mortality with low-dose computed tomographic screening,'' {\em New England Journal of Medicine}~{\bf 365}(5),  395--409 (2011).

\bibitem{deKoning2020}
de~Koning, H.~J. et~al., ``Reduced lung-cancer mortality with volume ct screening in a randomized trial,'' {\em New England Journal of Medicine}~{\bf 382}(6),  503--513 (2020).

\bibitem{Larici12017}
Larici, A.~R. et~al., ``Lung nodules: size still matters,'' {\em European Respiratory Review}~{\bf 26}(146) (2017).

\bibitem{Bankier2017}
Bankier, A. et~al., ``Recommendations for measuring pulmonary nodules at ct: A statement from the fleischner society,'' {\em Radiology}~{\bf 285}(2),  584--600 (2017).

\bibitem{Mozley2012}
Mozley, D. et~al., ``Measurement of tumor volumes improves recist-based response assessments in advanced lung cancer,'' {\em Translational Oncology}~{\bf 5}(1),  19--25 (2012).

\bibitem{Jaffe2006}
Jaffe, C.~C., ``Measures of response: Recist, who, and new alternatives,'' {\em Journal of Clinical Oncology}~{\bf 24}(20),  3245--3251 (2006).
\newblock PMID: 16829648.

\bibitem{Mahesh2023}
Mahesh, M. et~al., ``Patient exposure from radiologic and nuclear medicine procedures in the united states and worldwide: 2009-2018,'' {\em Radiology}~{\bf 301} (2023).

\bibitem{Leung2024}
Leung, K.~H., Rowe, S.~P., Sadaghiani, M.~S., Leal, J.~P., Mena, E., Choyke, P.~L., Du, Y., and Pomper, M.~G., ``Deep semisupervised transfer learning for fully automated whole-body tumor quantification and prognosis of cancer on pet/ct,'' {\em Journal of Nuclear Medicine}~{\bf 65}(4),  643--650 (2024).

\bibitem{Jennings2004}
Jennings, S.~G. et~al., ``Lung tumor growth: Assessment with ct—comparison of diameter and cross-sectional area with volume measurements,'' {\em Radiology}~{\bf 231}(3),  866--871 (2004).
\newblock PMID: 15163822.

\bibitem{Petrick2014}
Petrick, N. et~al., ``Comparison of 1d, 2d, and 3d nodule sizing methods by radiologists for spherical and complex nodules on thoracic ct phantom images,'' {\em Academic Radiology}~{\bf 21}(1),  30--40 (2014).

\bibitem{Lenz2023}
Brosch-Lenz, J., Uribe, C., Rahmim, A., and Saboury, B., ``Theranostic digital twins: An indispensable prerequisite for personalized cancer care,'' {\em Journal of Nuclear Medicine}~{\bf 64}(3),  501--501 (2023).

\bibitem{Tricarico2024}
Tricarico, P. et~al., ``Total metabolic tumor volume on 18f-fdg pet/ct is a game-changer for patients with metastatic lung cancer treated with immunotherapy,'' {\em Journal for ImmunoTherapy of Cancer}~{\bf 12}(4) (2024).

\bibitem{Liu2022}
Liu, J.~A., Yang, I.~Y., and Tsai, E.~B., ``Artificial intelligence (ai) for lung nodules, from the ajr special series on ai applications,'' {\em American Journal of Roentgenology}~{\bf 219}(5),  703--712 (2022).
\newblock PMID: 35544377.

\bibitem{Mellon2023}
{de Margerie-Mellon}, C. and Chassagnon, G., ``Artificial intelligence: A critical review of applications for lung nodule and lung cancer,'' {\em Diagnostic and Interventional Imaging}~{\bf 104}(1),  11--17 (2023).

\bibitem{Yoon2024}
Yoon, S.~H. et~al., ``Usefulness of longitudinal nodule-matching algorithm in computer-aided diagnosis of new pulmonary metastases on cancer surveillance ct scans,'' {\em Quantitative Imaging in Medicine and Surgery}~{\bf 14}(2) (2024).

\bibitem{Qi2020}
Qi, L.-L. et~al., ``Long-term follow-up of persistent pulmonary pure ground-glass nodules with deep learning–assisted nodule segmentation,'' {\em European Radiology}~{\bf 30}(1),  744–--755 (2020).

\bibitem{Xu2021}
Xu, Y. et~al., ``Consecutive serial non-contrast ct scan-based deep learning model facilitates the prediction of tumor invasiveness of ground-glass nodules,'' {\em Frontiers in Oncology}~{\bf 11} (2021).

\bibitem{Peters2023}
Peters, A.~A. et~al., ````will i change nodule management recommendations if i change my cad system?''—impact of volumetric deviation between different cad systems on lesion management,'' {\em Quantitative Imaging in Medicine and Surgery}~{\bf 33}(1),  5568–5577 (2023).

\bibitem{Li2019}
Li, L., Liu, Z., Huang, H., Lin, M., and Luo, D., ``Evaluating the performance of a deep learning-based computer-aided diagnosis (dl-cad) system for detecting and characterizing lung nodules: Comparison with the performance of double reading by radiologists,'' {\em Thoracic Cancer}~{\bf 10}(2),  183--192 (2019).

\bibitem{Hering2021}
Hering, A. et~al., ``Whole-body soft-tissue lesion tracking and segmentation in longitudinal {CT} imaging studies,'' in [{\em Proceedings of the Fourth Conference on Medical Imaging with Deep Learning}{\nolinebreak\hspace{0.1em}]},  {\em Proceedings of Machine Learning Research} {\bf 143},  312--326, PMLR (07--09 Jul 2021).

\bibitem{Chaudhry2022}
Chaudhry, H. A.~H. et~al., ``Unitochest: A lung image dataset for segmentation of cancerous nodules on ct scans,'' in [{\em Image Analysis and Processing -- ICIAP 2022}{\nolinebreak\hspace{0.1em}]},  Sclaroff, S., Distante, C., Leo, M., Farinella, G.~M., and Tombari, F., eds.,  185--196, Springer International Publishing, Cham (2022).

\bibitem{Wasserthal2023_TS}
Wasserthal, J. et~al., ``Totalsegmentator: Robust segmentation of 104 anatomic structures in ct images,'' {\em Radiology: Artificial Intelligence}~{\bf 5}(5),  e230024 (2023).

\bibitem{Isensee_2020}
Isensee, F. et~al., ``{nnU}-net: a self-configuring method for deep learning-based biomedical image segmentation,'' {\em Nature Methods}~{\bf 18}(2),  203--211 (2020).

\bibitem{Isensee2024}
Isensee, F. et~al., ``nnu-net revisited: A call for rigorous validation in 3d medical image segmentation,'' {\em ArXiv}~{\bf abs/2404.09556} (2024).

\bibitem{Zhuang2024}
Zhuang, Y. et~al., ``Mrisegmentator-abdomen: A fully automated multi-organ and structure segmentation tool for t1-weighted abdominal mri,'' (2024).

\bibitem{Zhou2023_nnFormer}
Zhou, H.-Y. et~al., ``nnformer: Volumetric medical image segmentation via a 3d transformer,'' {\em IEEE Transactions on Image Processing}~{\bf 32},  4036--4045 (2023).

\bibitem{drozdzal2016importance}
Drozdzal, M. et~al., ``The importance of skip connections in biomedical image segmentation,'' in [{\em International Workshop on Deep Learning in Medical Image Analysis, International Workshop on Large-Scale Annotation of Biomedical Data and Expert Label Synthesis}{\nolinebreak\hspace{0.1em}]},   179--187, Springer (2016).

\bibitem{Mathai2024}
Mathai, T., Liu, B., and Summers, R.~M., ``Segmentation of mediastinal lymph nodes in ct with anatomical priors,'' {\em International journal of computer assisted radiology and surgery}  (2024).

\bibitem{Bouget2022_StOlavs}
Bouget, D. et~al., ``Mediastinal lymph nodes segmentation using 3d convolutional neural network ensembles and anatomical priors guiding,'' {\em Computer Methods in Biomechanics and Biomedical Engineering: Imaging \& Visualization}~{\bf 11}(1),  44--58 (2023).

\bibitem{Kashyap2025}
Kashyap, M., Wang, X., Panjwani, N., Hasan, M., Zhang, Q., Huang, C., Bush, K., Chin, A., Vitzthum, L.~K., Dong, P., Zaky, S., Loo, B.~W., Diehn, M., Xing, L., Li, R., and Gensheimer, M.~F., ``Automated deep learning–based detection and segmentation of lung tumors at ct imaging,'' {\em Radiology}~{\bf 314}(1),  e233029 (2025).
\newblock PMID: 39835976.

\bibitem{Taupitz2007}
Taupitz, M., ``Imaging of lymph nodes -- mri and ct,'' {\em Springer} ,  321--329 (2007).

\bibitem{Jha2024}
Jha, A. et~al., ``Diagnostic performance of [68ga]dotatate pet/ct, [18f]fdg pet/ct, mri of the spine, and whole-body diagnostic ct and mri in the detection of spinal bone metastases associated with pheochromocytoma and paraganglioma,'' {\em European Radiology}  (2024).

\bibitem{Engelson2024}
Engelson, S., Ehrhardt, J., Kepp, T., Niemeijer, J., Schierholz, S., Berkel, L., Elser, Y., Sieren, M.~M., and Handels, H., ``{Comparison of anatomical priors for learning-based neural network guidance for mediastinal lymph node segmentation},'' in [{\em Medical Imaging 2024: Computer-Aided Diagnosis}{\nolinebreak\hspace{0.1em}]},  Chen, W. and Astley, S.~M., eds.,  {\bf 12927},  129271K, International Society for Optics and Photonics, SPIE (2024).

\bibitem{Ronneberger2015}
Ronneberger, O., Fischer, P., and Brox, T., ``U-net: Convolutional networks for biomedical image segmentation,'' in [{\em MICCAI 2015}{\nolinebreak\hspace{0.1em}]},   234--241, Springer International Publishing, Cham (2015).

\bibitem{Antonelli_2022}
Antonelli, M. et~al., ``The medical segmentation decathlon,'' {\em Nature Communications}~{\bf 13} (July 2022).

\bibitem{Lehmann2007}
Lehmann, G., ``Label object representation and manipulation with itk,'' {\em The Insight Journal}  (2007).

\end{thebibliography}
\bibliographystyle{spiebib} % makes bibtex use spiebib.bst

\clearpage

%% appendix 
\section{Appendix}
\label{sec_appendix}

\subsection{Model Details}

The base network is a UNet \cite{Isensee_2020,Ronneberger2015}, which was automatically optimized for the dataset based on its fingerprint. This included the determination of optimal hyper-parameters, such as a large batch size, adequate network depth among others. Other parameters, such as the number of epochs for training, were set to default values for training the 3D full-resolution nnUNet model. To optimize the network weights, stochastic gradient descent with Nesterov momentum ($\mu$ = 0.99), an initial learning rate of $10^{-3}$, and a batch size of 1 was utilized. For each output from the model, the corresponding ground truth segmentation mask was used for calculating the losses. A variety of data augmentations were implemented following the original implementation \cite{Isensee_2020}. All experiments were done on a workstation running Ubuntu 22.04 LTS with an NVIDIA Tesla V100 GPU. Experiments to determine the correct loss function to use were unnecessary as it was empirically found in previous investigations \cite{Isensee_2020,drozdzal2016importance} that the combination of the Dice and cross-entropy losses improved training stability and segmentation accuracy. The 3D ``\texttt{noPriors}'' and ``\texttt{withPriors}'' models took $\sim$2 and $\sim$3.5 days to complete training with 546 volumes respectively. Both models produced an output segmentation in $\sim$2 mins/volume.

\subsection{Statistical Analysis} 

The algorithms to compute detection metrics (precision, sensitivity, and F1-score) and segmentation metrics (Dice, Hausdorff Distance error) were obtained from the official Medical Segmentation Decathlon challenge \cite{Antonelli_2022}. The \texttt{LabelShapeStatisticsImageFilter} function \cite{Lehmann2007} in the \texttt{SimpleITK} package was utilized to compute the long- (Feret) and short-axis diameters, and the average of the two diameters was used to stratify nodules into clinically relevant nodules ($>$ 1cm) and those between 3 - 10 mm. A non-parametric Wilcoxon signed-rank test was performed to assess the utility of anatomical priors vs. no priors for lesion segmentation. A Friedman test was also conducted to determine any differences in performance across patient visits. A $p < 0.05$ indicated statistical significance. All statistical tests were performed using the ``wilcox.test'' and ``friedman.test'' functions from the ``stats'' package in RStudio (Version 2023.06.1+524) respectively.  

\subsection{Results - Additional Details}

\begin{figure}[!t]
    \centering
    \begin{minipage}{\linewidth}\centering    
    % \includegraphics[width=0.49\textwidth]{fig_results/bxp_segLungNodules_DiceAll.pdf}
    % \includegraphics[width=0.49\textwidth]{fig_results/bxp_segLungNodules_HDAll.pdf}
    % \includegraphics[width=0.49\textwidth]{fig_results/bxp_segLungNodules_DiceGreater.pdf}
    % \includegraphics[width=0.49\textwidth]{fig_results/bxp_segLungNodules_HDGreater.pdf}    
    % \subcaptionbox{Dice score}{\includegraphics[width=0.49\textwidth]
    % {fig_results/bxp_segLungNodules_DiceLesser.pdf}}
    % \subcaptionbox{Hausdorff Distance}{\includegraphics[width=0.49\textwidth]{fig_results/bxp_segLungNodules_HDLesser.pdf}}

    \includegraphics[width=0.325\textwidth]{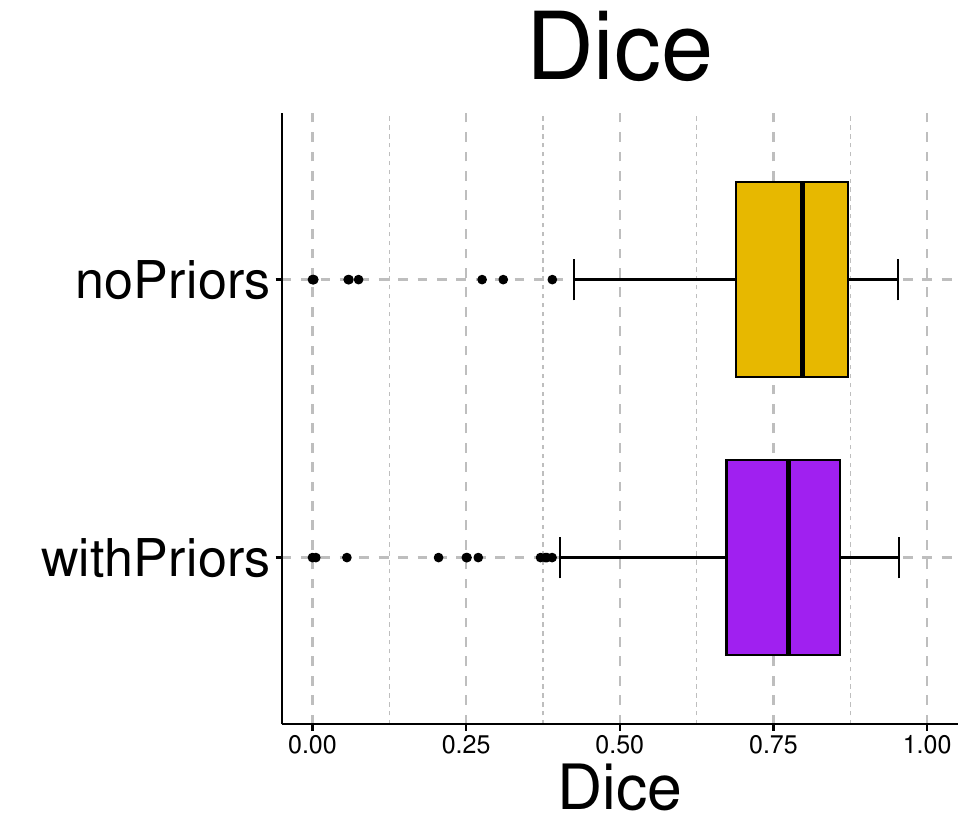}
    \includegraphics[width=0.325\textwidth]{fig_results/bxp_segLungNodules_DiceGreater.pdf}
    \includegraphics[width=0.325\textwidth]{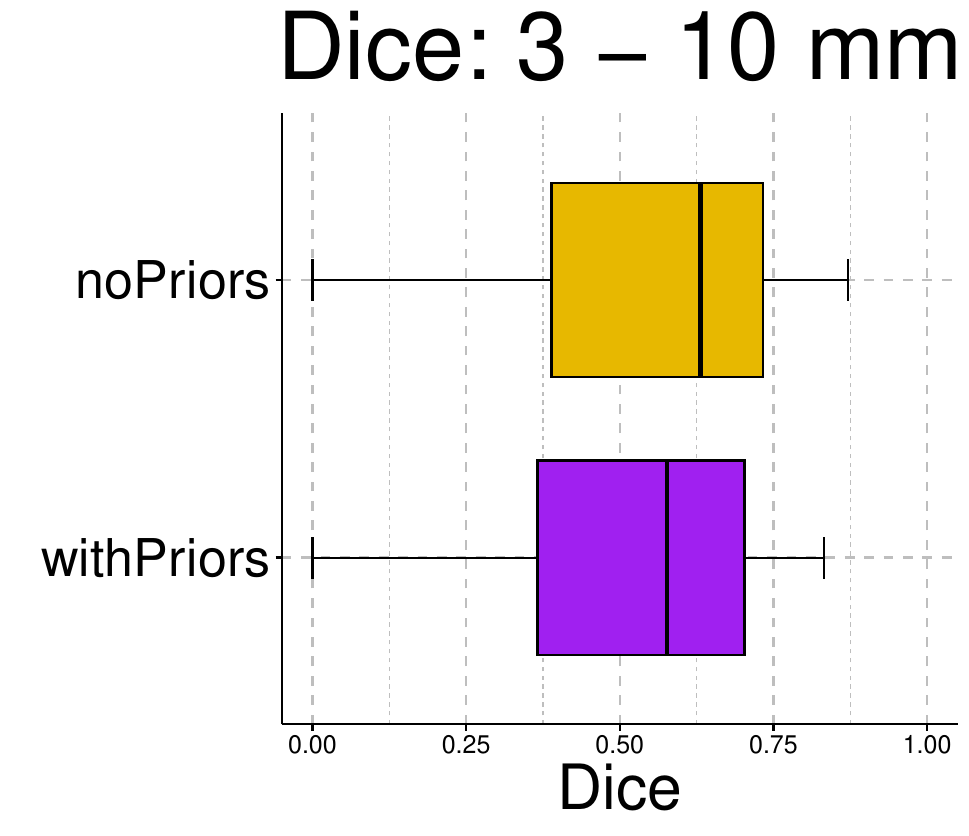}

    \medskip
    
    \includegraphics[width=0.325\textwidth]{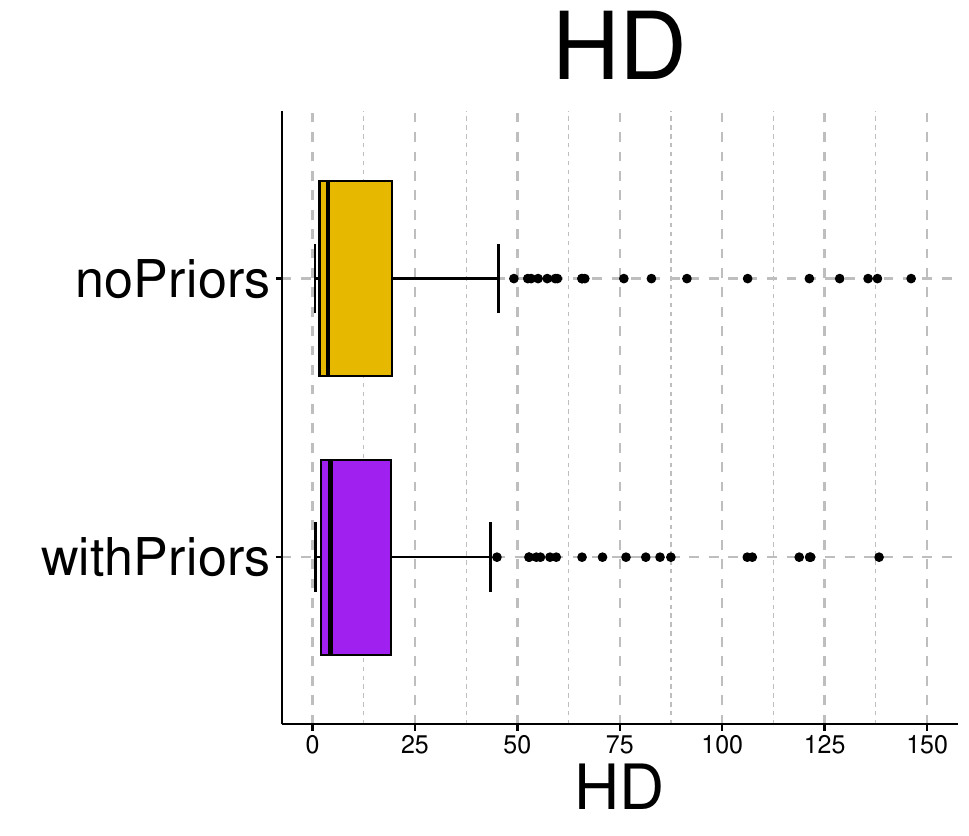}
    \includegraphics[width=0.325\textwidth]{fig_results/bxp_segLungNodules_HDGreater.pdf}
    \includegraphics[width=0.325\textwidth]{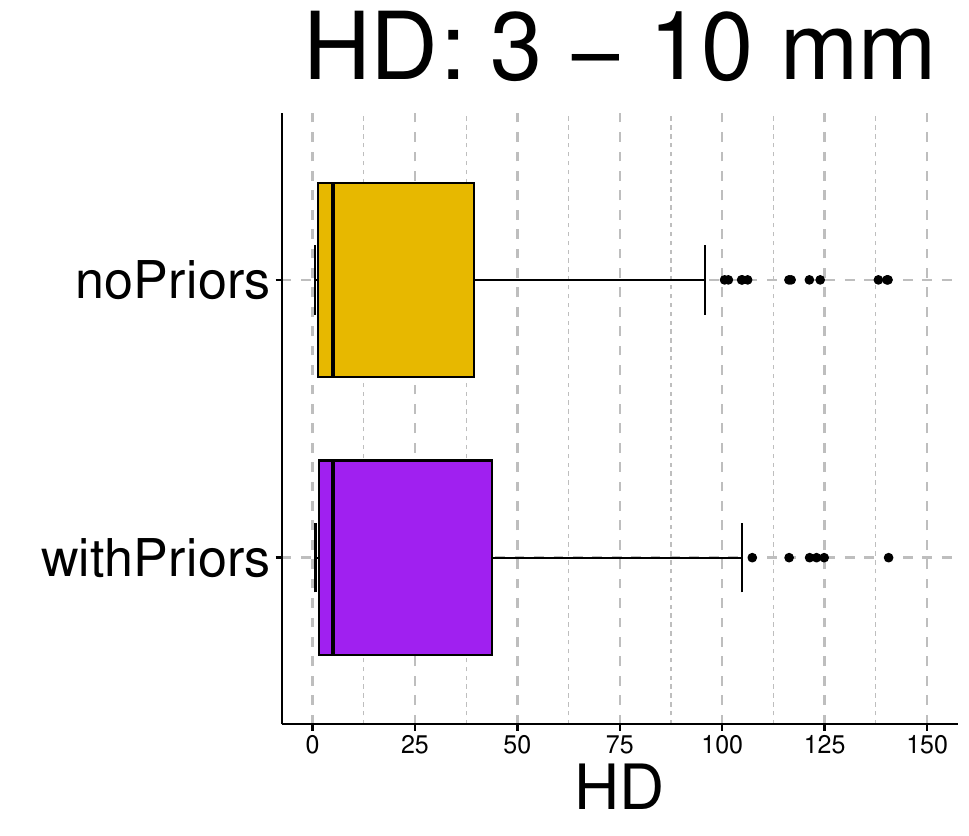}
       
    \caption{Dice scores (unitless, top row) and Hausdorff distance (HD) errors (mm, bottom row) for the segmentation of lung nodules of different sizes by the 3D full-resolution nnUNet models trained with (purple) and without (gold) anatomical priors.}
    
    \label{fig_results_boxplots}    
    \end{minipage}
\end{figure}

\begin{figure}[!t]
    \centering
    \begin{minipage}{\linewidth}\centering    
    \includegraphics[width=0.4\textwidth]{fig_results/r2_segLungNodules_noPriors.pdf}
    \includegraphics[width=0.4\textwidth]{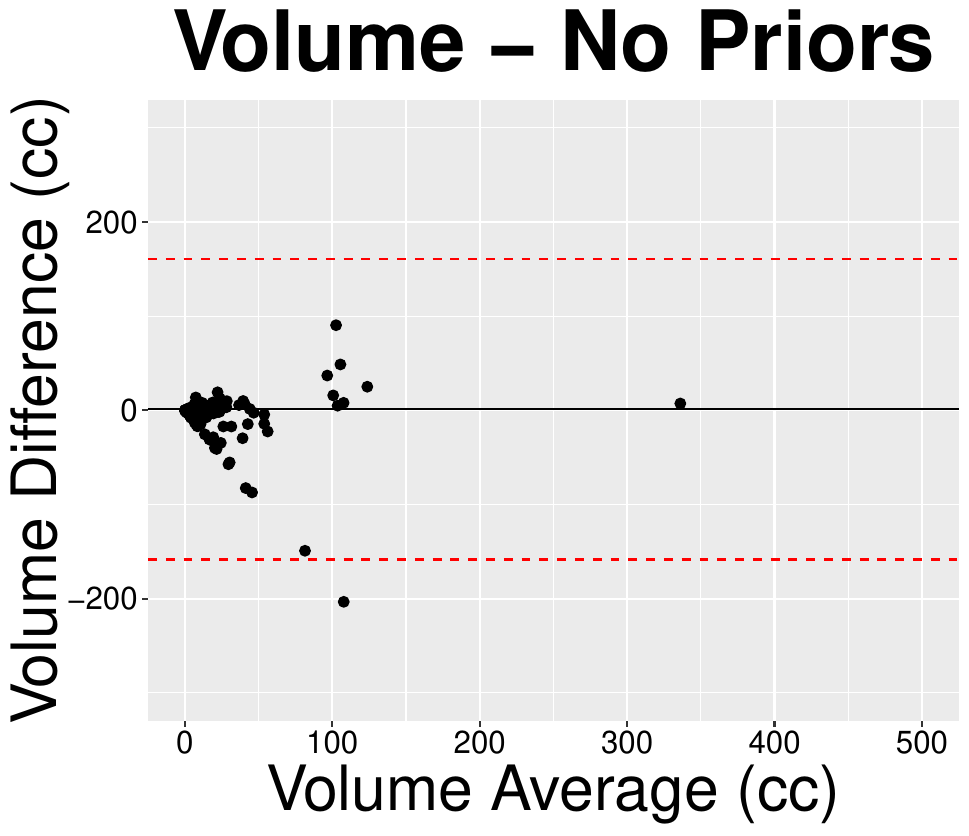}

    \medskip
    
    \subcaptionbox{Scatter Plot}{\includegraphics[width=0.4\textwidth]{fig_results/r2_segLungNodules_withPriors.pdf}}
    \subcaptionbox{Bland Altman Plot}{\includegraphics[width=0.4\textwidth]{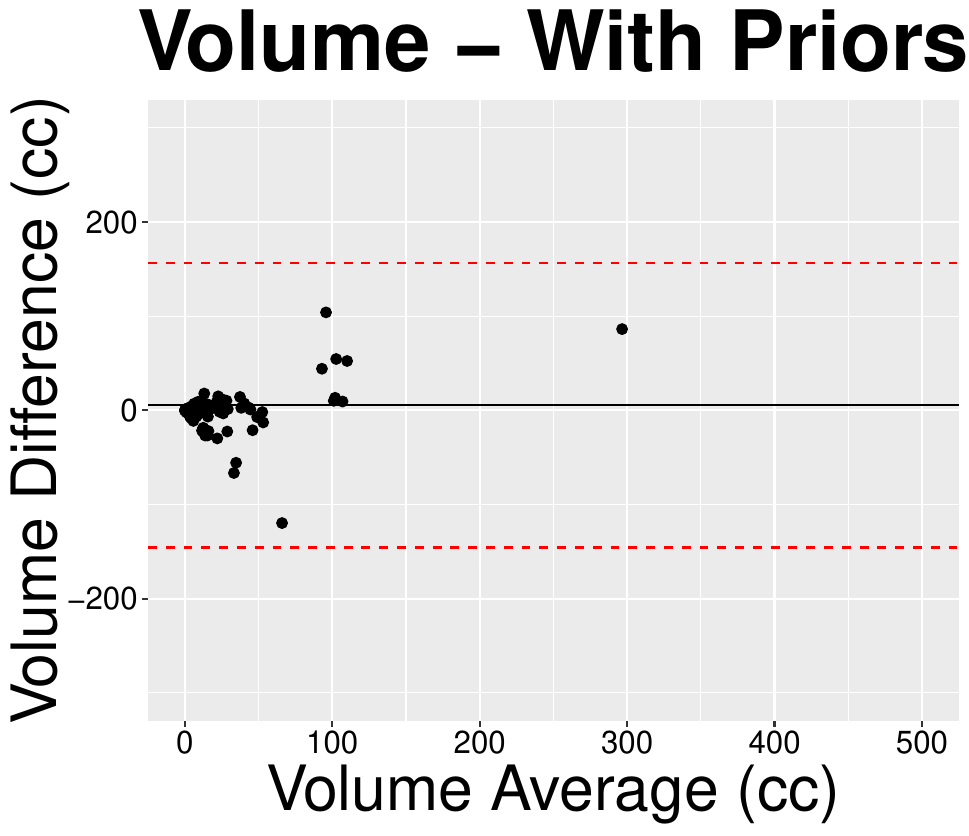}}
       
    \caption{Volumetric agreement between the ground-truth lung nodule annotations and the automated measurements. Column (a) shows the linear regression plots while column (b) shows the Bland-Altman plots. The gray bands in the scatter plots denote the 95\% confidence intervals (CI) for the best fit regression line (solid blue line). The upper and lower lines in column (b) represent the 95\% CI.}
    
    \label{fig_results_volumeAnalysis}    
    \end{minipage}
\end{figure}

Fig. \ref{fig_results_boxplots} shows box plots of the distribution of Dice and HD errors for different lesion sizes. The Dice scores were higher with the ``\texttt{noPriors}'' model, but the HD errors were lower with the ``\texttt{withPriors}'' model. Fig. \ref{fig_results_volumeAnalysis} shows the scatter and Bland-Altman (BA) plots that describe the agreement between the manual and automated measurements. A greater agreement was seen for the ``\texttt{withPriors}'' model ($R^{2} = .49$) in contrast to the ``\texttt{noPriors}'' model. The BA plots do not indicate any bias and show that both models tend to under-segment the total lesion volume, but the ``\texttt{withPriors}'' model showed a smaller spread of errors. 

\begin{figure}[!t]
    \centering
    \begin{subfigure}[b]{0.49\textwidth}
        \centering
        \includegraphics[width=\textwidth]{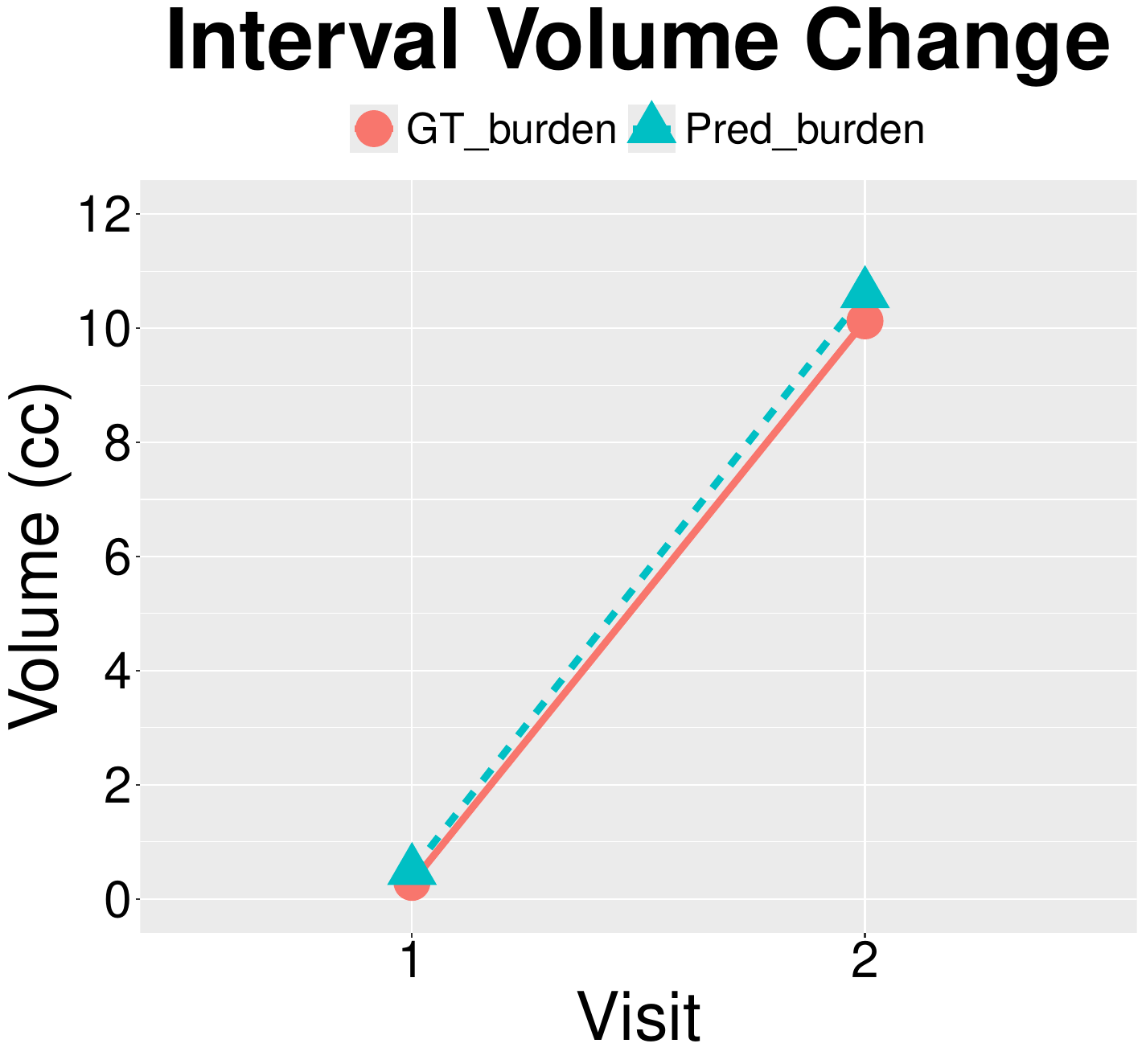}
        \caption{\small Patient 51}
    \end{subfigure}
    \hfill
    \begin{subfigure}[b]{0.49\textwidth}  
        \centering 
        \includegraphics[width=\textwidth]{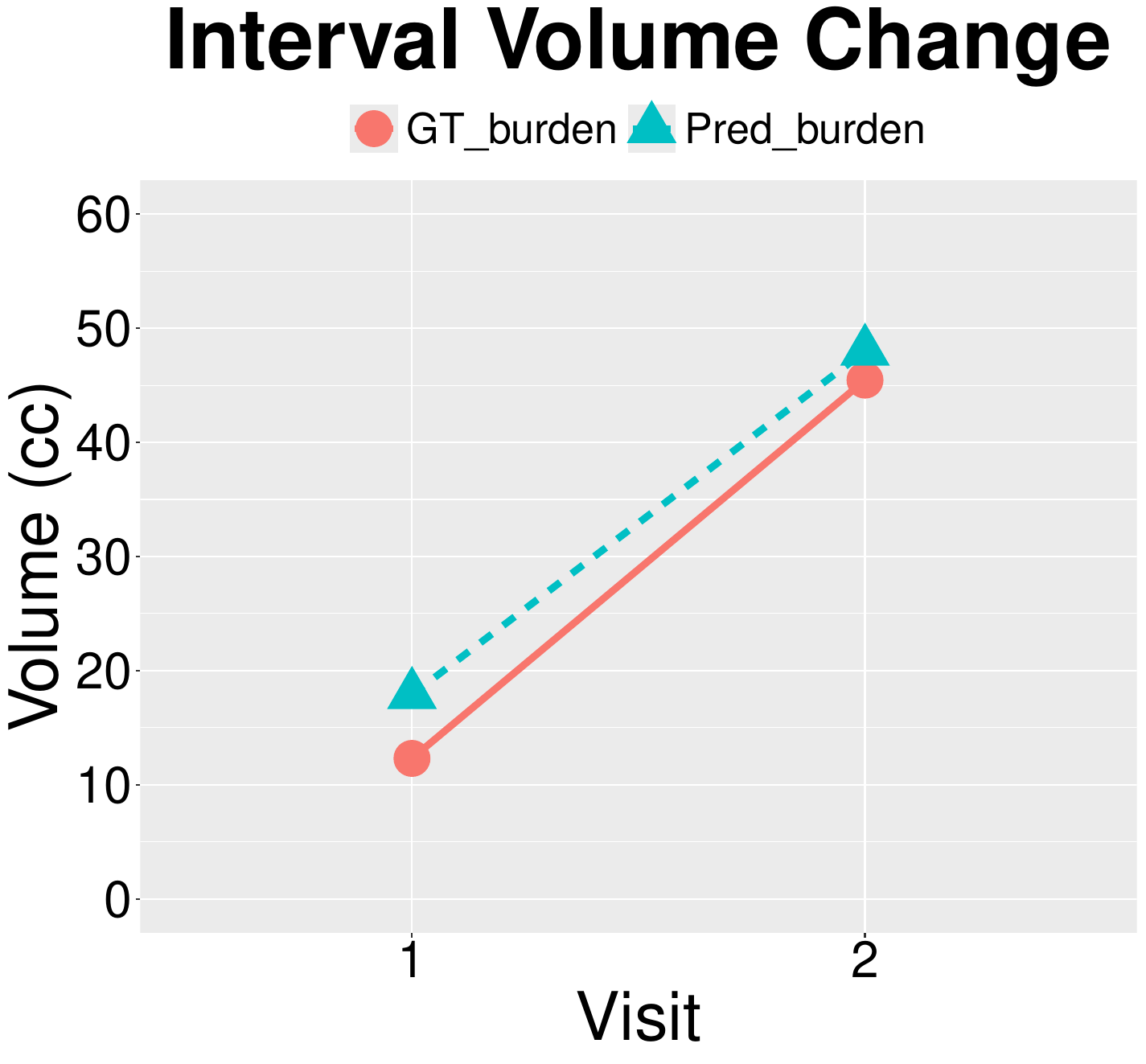}
        \caption{\small Patient 366}
    \end{subfigure}
    \vskip\baselineskip
    \begin{subfigure}[b]{0.49\textwidth}   
        \centering 
        \includegraphics[width=\textwidth]{fig_tumorBurden_intvChange/533_3visits.pdf}
        \caption{\small Patient 533}
    \end{subfigure}
    \hfill
    \begin{subfigure}[b]{0.49\textwidth}   
        \centering 
        \includegraphics[width=\textwidth]{fig_tumorBurden_intvChange/606_3visits.pdf}
        \caption{\small Patient 606}
    \end{subfigure}
    \caption{The trajectories of total lung tumor burden for four different patients computed from the segmentations of the 3D model trained with no anatomical priors. The ages at the time of the studies (initial and follow-up) were not provided in the UniToChest dataset. For patients 51 and 366 in (a) and (b) respectively, the total tumor burden increased over time. A dramatic increase in the tumor burden was seen especially for patient 366. In (c), the tumor burden for patient 533 at the first and last visits were greater than the ground truth annotation, and this was due to false positive predictions. Finally, for patient 606 in (d), the tumor burden rose steadily across the visits. } 
    \label{fig_tumorBurden_intervalChange}
\end{figure}

Fig. \ref{fig_tumorBurden_intervalChange} shows the manual vs. automated measurements for 4 separate patients with different lesion burdens across longitudinal exams. The median volume segmented by the ``\texttt{noPriors}'' model was 6.4 cc (IQR: 2.1, 18.1), while the median volume segmented by the ``\texttt{withPriors}'' model was 5.7 cc (IQR: 2.4, 16.5). The median volume difference (signed) between the manual and automated measurements for the ``\texttt{noPriors}'' model was 0.02 cc (IQR: -2.8, 1.2). The median volume difference (signed) for the ``\texttt{withPriors}'' model was -0.003 cc (IQR: -1.4, 1.4). 

%%%%%%%%%%%%%%%%%%%%%%%%%%%%%%%%%%%%%%%%%%%%%%%%%%%%%%%%%%%%%%%%%%
%%%%%%%%%%%%%%%%%%%%%%%%%%%%%%%%%%%%%%%%%%%%%%%%%%%%%%%%%%%%%%%%%%

%%%%%%%%%%%%%%%%%%%%%%%%%%%%%%%%%%%%%%%%%%%%%%%%%%%%%%%%%%%%%%%%%%
%%%%%%%%%%%%%%%%%%%%%%%%%%%%%%%%%%%%%%%%%%%%%%%%%%%%%%%%%%%%%%%%%%

%%%%%%%%%%%%%%%%%%%%%%%%%%%%%%%%%%%%%%%%%%%%%%%%%%%%%%%%%%%%%%%%%%
%%%%%%%%%%%%%%%%%%%%%%%%%%%%%%%%%%%%%%%%%%%%%%%%%%%%%%%%%%%%%%%%%%

\end{document}